\documentstyle[amssymb,twocolumn,aps,graphicx]{revtex}
\tighten
\begin{document}

\draft

\title{\bf Raman Response in Antiferromagnetic Two-Leg
$\mathbf{S=\frac{1}{2}}$ Heisenberg Ladders}

\author{Kai P. Schmidt, Christian Knetter and G\"otz S. Uhrig}

\address{Institut f\"ur Theoretische Physik, Universit\"at zu
  K\"oln, Z\"ulpicher Str. 77, D-50937 K\"oln, Germany\\[1mm]
  {\rm(\today)} }

\maketitle

\begin{abstract}
The Raman response in the antiferromagnetic 2-leg $S=1/2$
 Heisenberg ladder is calculated for various  couplings by
continuous unitary transformations.
For leg couplings above  80\% of the rung coupling a
characteristic 2-peak structure occurs
with a point of zero intensity within the continuum.
 Experimental data for ${\rm CaV_{2}O_{5}}$ and
${\rm La_{y}Ca_{14-y}Cu_{24}O_{41}}$ are analyzed and the coupling
constants are determined. Evidence is found that the Heisenberg model
is not sufficient to describe cuprate ladders.
We argue that a  cyclic exchange term is the appropriate extension.
\end{abstract}

\pacs{PACS numbers: 75.40.Gb, 75.50.Ee, 75.10.Jm}

\narrowtext

Strongly correlated electron systems in low dimensions are of
fundamental interest due to their fascinating properties resulting
from strong quantum fluctuations
\cite{Millis1,Sachdev1,Anderson1}. Important experimental insight
is gained from spectroscopic measurements of such systems. The
spectral densities measured yield information on the kinetics and
on the interaction of the elementary excitations as well as on the
matrix elements involved. Thus quantitative theoretical
calculations of spectral densities are a major task in condensed
matter physics. We use optimally chosen continuous unitary
transformations (CUT) to map complex
 many-body problems to a tractable few-body problems \cite{Uhrig1}. 
This clear
 concept serves as a perfect basis  to compute spectral densities of
strongly correlated systems thus establishing a quantitative contact
between theory and  experiment \cite{Uhrig2}.

We will focus on optical investigations, in particular on the Raman response,
of  antiferromagnetic 2-leg Heisenberg ladders realizing  quasi 
one-dimensional (1D) strongly correlated systems. There are several
experimental realizations of spin ladders like ${\rm CaV_{2}O_{5}}$,
SrCu$_2$O$_3$ and
${\rm La_{y}Ca_{14-y}Cu_{24}O_{41}}$
rendering  direct comparison between theory and experiment possible
\cite{Abrashev,Konstantinovic1,Popovic,Sugai1,Windt1}. 

Raman scattering measures
excitations with zero change of spin and  momentum. Starting at $T=0$
 from the $S=0$ ground state the singlet excitations at
zero momentum are probed. The Raman response in spin ladders was
recently calculated by first order perturbation theory for spin
ladders \cite{Jurecka1} and by exact diagonalization 
\cite{Suzuki1}. In this work, we present detailed predictions
obtained from CUTs using rung triplets as elementary
excitations. Our results are not resolution limited because
neither finite size effects occur nor  artificial broadenings are
necessary.

The Hamiltonian for the 2-leg Heisenberg ladder reads
\begin{equation}
  \label{H_start}
   H = \sum_{i}\left[ J_{\parallel}\left( {\bf S}_{1,i}{\bf S}_{1,i+1}+
{\bf S}_{2,i}{\bf S}_{2,i+1}\right) + J_{\perp}{\bf S}_{1,i}{\bf S}_{2,i}
 \right] \,
\end{equation}
where $J_{\parallel}>0$ and $J_{\perp}>0$ are the leg and rung couplings;
the subscript $i$ denotes the rungs and $1, 2$ the two legs.
At $T=0$ the Raman response $I(\omega)$ is given by the retarded resolvent
\begin{equation}
 \label{Intensity}
 I(\omega) = -\pi^{-1} {\rm Im}\left\langle0\left|
R^{\dagger}(\omega-H+i\delta)^{-1}R\right|0\right\rangle \ .
\end{equation}
The observables $R^{\rm rung}$ ($R^{\rm leg}$) for magnetic light
scattering in rung-rung (leg-leg) polarization read in leading order
\cite{Fleury1,Shastry1}
\begin{mathletters}
\label{Observables}
\begin{eqnarray}
 R^{\rm leg}& =& A_{0}^{\rm leg} \sum_{i}
\left( {\bf S}_{1,i}{\bf S}_{1,i+1} + {\bf S}_{2,i}{\bf S}_{2,i+1} \right) \\
 R^{\rm rung}& =& A^{\rm rung}_{0} \sum_{i} {\bf S}_{1,i}{\bf S}_{2,i} \ .
\end{eqnarray}
\end{mathletters}
The factors $A_{0}^{\rm leg}$ and $A^{\rm rung}_{0}$ depend on the
underlying
microscopic electronic model. It is beyond the scope of the present
work to compute them. Equally, we do not consider
resonating Raman excitation processes. 
Results will be given in units of the factors squared.

Technically,  we employ a CUT to map the
Hamiltonian $H$ to an effective Hamiltonian $H_{\rm eff}$ which conserves the
number of rung-triplets, i.e.\ $[H_{\rm 0},H_{\rm eff}]=0$ where
$H_{\rm 0}:=H|_{J_{\rm \parallel}=0}$ \cite{Uhrig1,Uhrig2,Knetter-neu}.
The ground state of $H_{\rm eff}$ is the rung-triplet vacuum.
For the response function $I(\omega)$ the observable $R$ is mapped by
 the same unitary transformation
 as the Hamiltonian to an
effective observable $R_{\rm eff}$.
We implemented the CUT  perturbatively 
in $x:=J_{\parallel}/J_{\perp}$ and  calculated
 $H_{\rm eff}$  to high orders (1-triplet terms: 14$^{\rm th}$, 
2-triplet terms: 13$^{\rm th}$ order). The effective
observable $R_{\rm eff}$ is computed
to order $10$ in the 2-triplet sector and to order 7 in the
4-triplet sector. Generally, higher orders make higher accuracy
possible. As a rule of thumb, a calculation in order $n$ accounts for
hopping  or interaction processes 
extending over a distance of $n$ rungs.

The truncated series gives quantitative results up to $x\approx 0.6$.
Using standard extrapolation techniques like Pad\'e approximants and optimized
 perturbation theory \cite{Domb,Kleinert},
the effective operators $R_{\rm eff}$ and $H_{\rm eff}$
can be calculated up to $x\approx 1$ with an
uncertainty of about $5\%$. A qualitative
description is obtained for $x\approx 1.2$. The Raman spectral density is
 calculated as continued fraction by
 tridiagonalization \cite{Gagliano1}. Because in 1D
 asymptotically free particles with quadratic dispersion
display square root behavior at the band edges (van-Hove singularities)
 we use a square root terminator for the continued fraction.
Thus neither finite size nor finite resolution  affects our results.
{\em No} divergences
occur because a rung cannot be excited twice,
 i.e.\ triplets exclude one another. The
relative motion of a pair of triplets corresponds to the dynamics of
a particle on a half-infinite chain \cite{Uhrig3Uhrig4}.

\begin{figure}[htbp]
  \begin{center}
    \includegraphics[width=8.2cm]{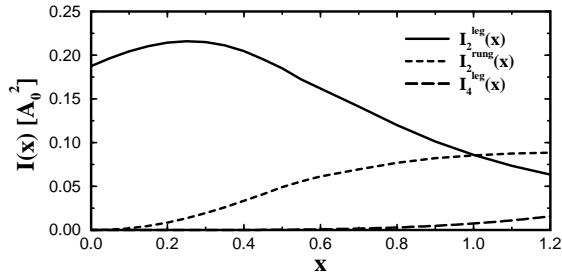}
    \caption{Spectral weights $I_n$ as function of
$x=J_\parallel/J_\perp$; $n$ denotes the number of triplets involved;
the superscript denote the observable as defined in Eq.~(\ref{Observables}).}
    \label{fig:Raman_Itot}
  \end{center}
\end{figure}

Sectors with odd number of triplets are inaccessible by
Raman scattering due to the invariance of the two observables
$R^{\rm leg}_{\rm eff}$ and
 $R^{\rm rung}_{\rm eff}$ with respect to reflection about the centerline
of the ladder.  Thus only  excitations with even number of
triplets matter. Therefore the leading contributions to the Raman
response come from the 2-triplet and the 4-triplet sector. The
total spectral weight (integrated over  frequencies and momenta)
 of these contributions is depicted in Fig.~\ref{fig:Raman_Itot}
 \cite{note}.
 The 4-triplet
sector has a  spectral weight $I_4$ of less than $7\%$ of the spectral weight
$I_2$  of the 2-triplet sector
 at $x \approx 1$. Hence we focus on the 2-triplet contribution in the
sequel.

\begin{figure}[htbp]
  \begin{center}
    \includegraphics[width=8.2cm]{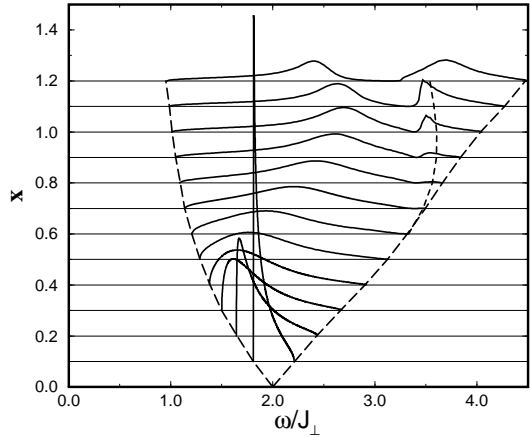}
    \caption{Raman spectral density $I(\omega)$
      for various coupling ratios $x$. For each curve the total weight is
      set to
      $0.1 [{\rm A_{0}^{2}}]$. Long dashed lines depict  lower and upper band
      edge of the 2-triplet continuum. The short dashed line shows
      $2\omega(k=0)$ for $x > 0.5$ where this is {\em not} the upper band edge
      ($\omega(k)$ is the 1-triplet dispersion, see
      inset in Fig.~\ref{fig:Sugai}).}
    \label{fig:Raman_Film}
  \end{center}
\end{figure}

In Fig.~\ref{fig:Raman_Film} the spectral density $I(\omega)$ from the
2-triplet sector is shown for various values of $x=J_\parallel/J_\perp$.
 The line {\em shape} is the {\em same} for
$R^{\rm rung}$ and $R^{\rm leg}$ \cite{Singh1} because the  Hamiltonian is
a weighted sum of the two observables
$H = R^{\rm rung}+xR^{\rm leg}$ (for $A_0=1$). Thus the excited state
$R^{\rm rung}|0\rangle$ equals $-xR^{\rm leg}|0\rangle$ except for
a component proportional to the ground state $|0\rangle$ which does not
matter at finite frequencies. This fact leads also to the intersection
at $x=1$ visible in Fig.\ref{fig:Raman_Itot}.

The spread of the lines in Fig.~2 on increasing $x$ indicates clearly the
increasing band width.  For small $x$ the Raman intensity shows a strong
resonance near the lower band edge.
 This resonance is a consequence of the 2-triplet attraction on neighboring
sites \cite{Uhrig3Uhrig4,Jurecka1}. Above a certain finite
 total momentum this leads to a 2-triplet bound state
\cite{Uhrig3Uhrig4,Sushkov1,Sachdev2,Trebst1,Zheng1,Brenig1,Windt1}
of which the resonance at the lower band edge is a precursor.
Since  for larger
 $x$ the kinetic energy of the relative motion of the triplets increases
 the influence of the attraction decreases. Therefore, the  resonance is
rapidly broadened and shifted to the center of the continuum.
In view of analyses  of the spin gap
\cite{Sugai1} we note that it is not possible to detect the onset
of the 2-triplet, non-resonant 
Raman continuum, 
i.e.\ twice the spin gap, reliably for  $x\gtrapprox 0.4$. 
Furthermore, we found that the non-resonant line shapes do not depend
very much on the precise form of the excited state $R|0\rangle$.
The qualitative features depend more on kinetics
and  on interaction.

In Fig.~2 for $x>0.6$, a second peak is visible near the upper boundary of the
2-triplet continuum, becoming more pronounced
on increasing $x$. This feature is the  combined effect of 1-triplet
kinetics, 2-triplet interaction and matrix elements.
First, the occurrence of a  dip in the 1-triplet dispersion $\omega(k)$
at $k=0$ (cf.\ inset in Fig.~\ref{fig:Sugai}) leads to  an additional
 van-Hove singularity situated at $2\omega(0)$ providing
additional spectral weight. The importance of this effect is
illustrated by the short dashed line in Fig.~\ref{fig:Raman_Film}.
Second, the additional spectral weight is separated from the
main peak by a double zero in the spectral density, see
Fig.~\ref{fig:Raman_Film}. This double zero stems from a simple zero
in the matrix elements implying that at a certain energy $\omega$
the state $R|0\rangle$ is orthogonal to the excited state $|\omega\rangle$ !
This intriguing phenomenon results from destructive interference
between several coupling contributions. We found that the destructive
interference is triggered by the 2-triplet interaction since it
vanishes when the 2-triplet interaction is switched off by hand
\cite{Knetter1}. Hence the orthogonality is induced by the interaction
recalling in a broad sense an orthogonality catastrophe.
Indeed, an arbitrarily small amount of the interaction
suffices  to induce at least a very narrow dip with the double zero
at its bottom. We are led
to the conclusion that the large density of states provided by the
additional van-Hove singularity renders the system particularly susceptible
to the interaction-induced orthogonality.

\begin{figure}[htbp]
  \begin{center}
    \includegraphics[width=8.2cm]{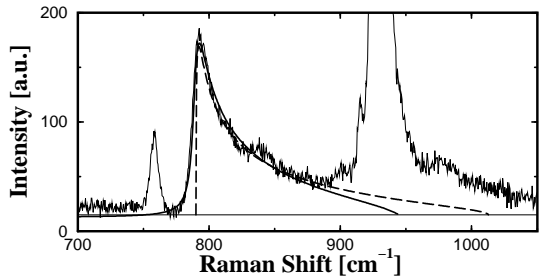}
    \caption{Raman response for CaV$_{2}$O$_{5}$ from
      Ref.~\protect\cite{Konstantinovic1}.
      Solid line: fit with $x=0.09,
      J_{\rm \perp}=431 {\rm cm^{-1}}$
      assuming an experimental resolution of $\Gamma = 3 {\rm cm}^{-1}$.
      Long dashed line: fit with $x=0.125,
      J_{\rm \perp}=447 {\rm cm^{-1}}, \Gamma=0$.
      Thin constant line: offset of the fits to account for background.}
    \label{fig:Cavo}
  \end{center}
\end{figure}

\begin{figure}[htbp]
  \begin{center}
    \includegraphics[width=8.2cm]{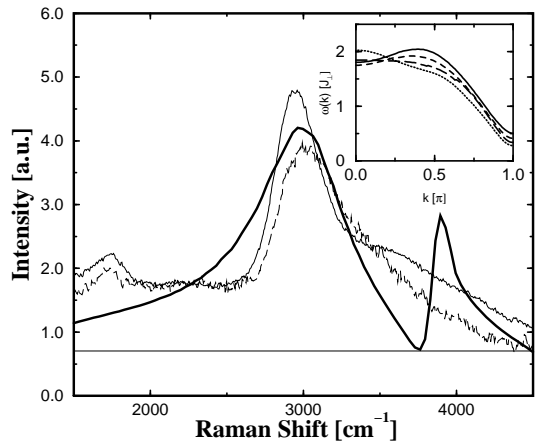}
    \caption{Raman response for ${\rm La_{6}Ca_{8}Cu_{24}O_{41}}$. Rough thin
      lines: experimental data \protect\cite{Sugai1} in $(aa)$ (dashed)
      and $(cc)$ (solid)
      polarization scaled to the same constant value between 2000 and 2500
      cm$^{-1}$ . Thick solid line:  theory for
      $J_\parallel=J_\perp=1100{\rm cm}^{-1}$ and
      resolution $\Gamma=3{\rm cm^{-1}}$.  Thin constant line: offset of the
      fit to account for background. Inset: 1-triplet
      dispersions $\omega(k)$ for $J_\parallel=J_\perp$
      and $x_{\rm cyc}=0, 0.05, 0.10, 0.15$
      (solid, dashed, long dashed, dotted).}
    \label{fig:Sugai}
  \end{center}
\end{figure}

Analyzing experimental data,
in Fig.~\ref{fig:Cavo} the Raman shift for ${\rm CaV_{2}O_{5}}$ 
 \cite{Konstantinovic1} is shown. This substance
 is a quasi-2D layered material where the $S=1/2$ $\rm V^{4+}$ ions
 form weakly coupled 2-leg ladders (trellis lattice).
Susceptibility measurements \cite{Johnston1}
 predict weakly interacting rungs with $x \approx 0.1$.
Theoretical analysis in first order yields
$x=0.11$ and $J=447 {\rm cm^{-1}}$ \cite{Jurecka1}. For these values
of $x$ our results can be considered to be exact.
It is instructive to compare fits based on our results to
the previous analysis. Two fits and their parameters are given in
Fig.~\ref{fig:Cavo}. Assuming the same experimental
resolution as  in Ref.~\onlinecite{Jurecka1} we find $x=0.09$ deviating
by 20\% from the value obtained in first order. 
A change $\Delta x$ of order $x^2$ was to be expected
since terms of order $x^2$
were neglected in the first order analysis \cite{Jurecka1}. Hence the
 relative change  $\Delta x/x$ is of the order of $x$.
 We have to stress that the fits are
extremely sensitive to the experimental resolution assumed. Assuming zero
resolution ($\Gamma=0$, see Fig.~\ref{fig:Cavo})
the best value of $x$ is $0.125$, i.e.\ it is changed by about
40\%. This remarkable sensitivity results from the very narrow
resonance the height of which changes quickly as function of the resolution.
An effective value of $\Gamma$ adding to the experimental resolution
might be generated by
residual couplings not considered explicitly.
It is unfortunate that the dominant phonon at 940cm$^{-1}$ prevents
the observation of the upper part of the continuum
so that the theoretically possible high precision analysis cannot be
performed. Within the non-resonant Raman theory, our results 
exclude that the shoulder at 
$970 {\rm cm}^{-1}$ besides the dominant phonon is of 
magnetic origin. We conclude that the previous analyses
\cite{Johnston1,Konstantinovic1,Suzuki1,Jurecka1} and ours
 agree within the achievable accuracy.

In Fig.~\ref{fig:Sugai} the Raman lines \cite{Sugai1} for
${\rm La_{6}Ca_{8}Cu_{24}O_{41}}$ are analysed.
${\rm La_{6}Ca_{8}Cu_{24}O_{41}}$ is a layered material containing
${\rm CuO_{2}}$ 1D spin chains and ${\rm Cu_{2}O_{3}}$ 2-leg
spin ladders \cite{Carron1}. The inter-ladder coupling is weak and
frustrated (trellis lattice) so that the ladders can be treated as
isolated ladders. Because the atomic distance between
neighboring copper sites in rung and leg direction is almost the same one
expects the spin ladders to be in the isotropic regime $x\approx 1$.
This view is corroborated by the analysis of the 2-triplet bound
state observed by infrared (IR) absorption \cite{Windt1} leading to
values of $J_{\perp}$ between  1020 and 1100 ${\rm cm^{-1}}$.
 On the other hand, the spin gap values and  neutron scattering data
cannot be reconciled with the model in Eq.~\ref{H_start} for
$J_\parallel\approx J_\perp$, see e.g.\ \cite{Brehmer1} and references
therein. Either  values of $x$  beyond unity or extensions of the model
are necessary.

Assuming the model in Eq.~\ref{H_start} and the observables in 
Eq.~\ref{Observables} the  line shapes for
the two polarizations  should be identical independent of the
value of $x$ as explained above. The fact that this is not the
case, see Fig.~\ref{fig:Sugai},  indicates that
 an extension of the  model is  necessary though it cannot be excluded that a
modification of the observables (\ref{Observables}) would also explain the
 deviations.

The $x=1$ result  agrees quite well with the experimental data.
The value $J_\perp=1100{\rm cm}^{-1}$ is consistent with the
IR result \cite{Windt1}. The main peak
is situated at the right energy and has approximately the right width.
 The weight of the experimental high energy shoulder corresponds
to the weight in the theoretical second peak at higher energies.
However, important deviations remain.  (i) Experimentally,
there is a high energy shoulder but no second peak.
(ii) The main experimental peak is
 sharper.
Inspecting Fig.~\ref{fig:Raman_Film}  these
discrepancies are not remedied by assuming a larger coupling ratio $x$.
Furthermore, the sensitivity on the excited vector $R|0\rangle$ is
not very large so that an explanation of the deviations in terms
of modifications of the observables $R$ in (\ref{Observables}) is
unlikely. Hence extensions of the model must indeed be
considered.

Numerous results favor the inclusion of a
4-spin cyclic exchange term $H_{\rm cyc}$ as next important term
\cite{Schmidt1,Honda1,Reischl1,Coldea1}. Especially in the spin ladder
material ${\rm La_{6}Ca_{8}Cu_{24}O_{41}}$  about 10\%
cyclic exchange reconcile results for the spin gap, the  dispersion and the
weighted spectral densities \cite{Brehmer1,Eccleston1,Windt1}.
The term $H_{\rm cyc}$ reads
\begin{eqnarray}
&& H_{\rm cyc} =  2 J_{\rm cyc}\sum_i
 [ ({\bf S}_{1,i}{\bf S}_{1,i+1})({\bf S}_{2,i}{\bf S}_{2,i+1}) + \\
 & &  ({\bf S}_{1,i}{\bf S}_{2,i})({\bf S}_{1,i+1}{\bf S}_{2,i+1}) -
({\bf S}_{1,i}{\bf S}_{2,i+1})({\bf S}_{1,i+1}{\bf S}_{2,i})  ] \
. \nonumber
 \end{eqnarray}
To assess the effect of a cyclic exchange term on the Raman line
shape qualitatively we include the 1-triplet dispersion
for various values of $x_{\rm cyc}=J_{\rm cyc}/J_{\rm perp}$
in the inset in Fig.~\ref{fig:Sugai}.
The gap $\omega(\pi)$ decreases on increasing
$x_{\rm cyc}$ \cite{Brehmer1}. For $x_{\rm cyc}=0.1$ we find
$\Delta=0.337 J_\perp$ in very good agreement with the  result from
neutron scattering $\Delta=0.343 J_\perp$ \cite{Eccleston1}.
More important for the Raman line shape is that the
dip in the dispersion $\omega(k)$ at $k=0$ is reduced
 on increasing
$x_{\rm cyc}$. The dispersion is monotonic for $x_{\rm cyc}=0.10$.
Additionally, preliminary results on the 2-triplet terms show that
the attractive interaction between triplets is diminished by
the cyclic exchange. Thus we expect the pronounced zero inside the continuum
of the Raman line to disappear on inclusion of cyclic exchange
leading to an asymmetric broad peak with a shoulder at the
 high energy side as measured, see Fig.~\ref{fig:Sugai} and
Refs.~\onlinecite{Abrashev,Popovic,Sugai1}. We estimate that the deviation
(i) is remedied by a 4-spin cyclic exchange term with $x_{\rm
cyc}\approx 0.1$ in agreement with other analyses
\cite{Brehmer1,Eccleston1,Windt1}.

The deviation (ii) in the sharpness of the main peak may
also be  remedied by the inclusion of the 4-spin cyclic
exchange term. Alternatively, one might invoke a doping effect,
namely holes in the ladder, since the lines are considerably
sharper for Sr$_{14}$Cu$_{24}$O$_{41}$ \cite{Sugai1}, in particular
in ($cc$) polarization. Other results \cite{Windt1,nucke00},
however, show that there are no holes in the ladders
shedding doubt on an explanation in terms of holes in the ladders.

In conclusion, we calculated the Raman response in the antiferromagnetic
 2-leg $S=1/2$ Heisenberg ladder in terms  of
 dressed rung-triplets as elementary excitations. Our results
are based on a continuous unitary transformation introducing the number
of elementary excitations as good quantum number. We demonstrated that
the 2-triplet contributions dominate by far.
Unexpectedly we found a 2-peak structure composed of a broad main peak
and a secondary peak at higher energies for $J_\parallel \gtrapprox 0.8
J_\perp$. From the inconsistency with the experimental finding for
${\rm La_{6}Ca_{8}Cu_{24}O_{41}}$ we infer that the  model must
 be extended for which a 4-spin cyclic exchange term is the best candidate
to date. To exclude possible effects due to doping we urge for further
investigations in undoped materials such as
${\rm SrCu_{2}O_{3}}$.

We gratefully acknowledge M.\ Gr\"uninger and
E.\ M\"uller-Hartmann for very helpful discussions and
 M.\ Konstantinovi\'c for  kind provision of
experimental data. This work is supported by the
DFG in SP1073.

\end{document}